# An Innovative Workspace for The Cherenkov Telescope Array

Alessandro Costa *, Eva Sciacca *, Ugo Becciani *, Piero Massimino *,
Simone Riggi *, David Sanchez † and Fabio Vitello *
* INAF-Osservatorio Astrofisico di Catania, Italy
† Laboratoire d'Annecy-le-Vieux de Physique des Particules, France
Email: alessandro.costa@oact.inaf.it

*Abstract*—The Cherenkov Telescope Array (CTA) is an initiative to build the next generation, ground-based gamma-ray observatories. We present a prototype workspace developed at INAF that aims at providing innovative solutions for the CTA community. The workspace leverages open source technologies providing web access to a set of tools widely used by the CTA community. Two different user interaction models, connected to an authentication and authorization infrastructure, have been implemented in this workspace. The first one is a workflow management system accessed via a science gateway (based on the Liferay platform) and the second one is an interactive virtual desktop environment. The integrated workflow system allows to run applications used in astronomy and physics researches into distributed computing infrastructures (ranging from clusters to grids and clouds). The interactive desktop environment allows to use many software packages without any installation on local desktops exploiting their native graphical user interfaces. The science gateway and the interactive desktop environment are connected to the authentication and authorization infrastructure composed by a Shibboleth identity provider and a Grouper authorization solution. The Grouper released attributes are consumed by the science gateway to authorize the access to specific web resources and the role management mechanism in Liferay provides the attribute-role mapping.

## I. INTRODUCTION

The Cherenkov Telescope Array (CTA) project[1] aims at building a new observatory for very high-energy (VHE) gamma rays [1]. CTA has ambitious science goals focused in understanding the origin of cosmic rays and their role in the Universe, the nature and variety of particle acceleration around black holes and in searching for the ultimate nature of matter and physics beyond the Standard Model. For reaching these goals it is aimed to achieve full-sky coverage by deploying hundreds of telescopes at two sites in the southern and the northern hemispheres.

To guarantee the smooth running of the complex CTA observatory three main management elements have been identified: (i) the Science Operation Centre, which is in charge of the organisation of observations, (ii) the Array Operation Centre, which conducts the operation, monitors the telescopes and the atmosphere, and provides all calibration and environmental data necessary for the analysis, and (iii) the Science Data Centre, which provides and disseminates data and analysis software to the science community at large, using common astronomical standards and existing computing infrastructures.

The total data volume to be managed by the CTA Science Data Centre is of the order of 27 PB/year [2], when all dataset versions and backup replicas are considered. All levels of data (from the raw data to the high-level final products) will be archived in a standardised way, to allow access and reprocessing. The "CTA science gateway" will provide access to data, support services, software and data center infrastructures. It is foreseen that individual scientists using the analysis software made available by CTA can conduct the high-level analysis of CTA data. The Gateway aims at supporting workflow handling, virtualization of hardware, visualization as well as resource discovery, job execution, access to data collections, and applications and tools for data analysis.

Access to the developed services within the CTA science gateway and other CTA web resources will be based on each users profile and category (e.g. unsigned user, guest observer, advanced user, principal investigator, archive user, pipeline user, etc). For such a purpose the Authentication and Authorization infrastructure (AAI) plays a key role in the scientific process and will be widely discussed in this paper. The AAI is also a fundamental part of the workspace developed by INAF and described in this work.

The "CTA science gateway" is implemented as a set of complementary modules. Three of them are being developed with different aims: the first one is developed by INAF and presented in this paper, it provides a workflow management system, it is powered by WS-PGRADE/gUSE[2] [3], based on Liferay platform [3] and an added value of this module is a web-desktop environment; the second one integrates existing CTA applications in a specific InSilicoLab platform [4] developed by Cyfronet; and the third module, developed by the Observatoire de Paris, is compliant with the Virtual Observatory and it is based on the Django platform.

The INAF workspace is composed by a science gateway module and by the Authentication and Authorization Infrastructure. The science gateway, first introduced in [5], allows the user to access a workflow management system with a customizable graphical web user interface (see Section II-A) and a web-desktop environment (see Section III). The gateway has been further developed to enable the processing of the Fermi Workflow Demonstrator (see Section II-B) and connected to the Authentication and Authorization infrastructure developed by INAF (see Section IV). Moreover the gateway has been combined and integrated with the other modules put in place by the Observatoire de Paris and Cyfronet using a common menu-bar and the shared authentication and authorization infrastructure.

---

[1] CTA project web page: https://portal.cta-observatory.org
[2] WS-PGRADE/gUSE web page: http://guse.hu
[3] Liferay web page: https://www.liferay.com

## II. THE INAF CTA SCIENCE GATEWAY

The INAF CTA science gateway (available at the following URL: http://cta-sg.oact.inaf.it/ ) is aiming at providing a web instrument for high energy astrophysics. It leverages on open source technologies giving web access to a set of tools and software widely used by the CTA community. An extended (though not exhaustive) list[4] of tools provided by this technology embrace XANADU software package, GammaLib & ctools, Fermi Science Tools, Aladin, IRAF. Each tool is available interactively via a dedicated web-desktop environment or through a workflow management system.

The gateway is based on the Liferay platform. Liferay is an enterprise-level framework, offering both an advanced development infrastructure and a flexible content management system. We used the Liferay community edition that is released under an open source GNU LGPL license. This provides a cutting-edge and inexpensive solution that best suits our purposes; Liferay is moreover used by a wide community of users. These aspects are an important added value for a technology destined to follow the CTA consortium for its lifetime. Liferay platform has a large set of configurations implementing High Availability (HA) solutions. The resulting Liferay system will be able to handle the expected number of concurrent users and subsequent traffic, and will reduce single points of failure resulting in a more robust system. Liferay can also be configured to load balancing and clustering at the server level. Both user profile management and workspace applications are provided by the Liferay platform and can be easily improved and customized according to the CTA present and future requirements. INAF CTA science gateway provides a workflow management system with a customizable graphical web user interface [6] and a web-desktop environment.

The integrated workflow system (based on gUSE/WS-PGRADE) seamlessly enables the execution of astronomical and physics workflows (and jobs) on major platforms such as DIRAC INTERWARE [5], ARC [6], Globus[7], gLite [8], UNICORE [9], PBS [10] as well as web services or clouds [7].

The web-desktop environment: Astronomical & Physics Cloud Interactive Desktop (ACID), allows to use many software packages without any installation on local desktops exploiting their native Graphical User Interface. Finally, a common menu bar has been added to allow the integration with the different CTA science gateway modules. The role of the common menu-bar is to i) standardize the layout of each module using a shared Common cascading Style Sheet; ii) prompt the user with a common top-menu that provides access to each module; iii) prompt the user with information about her/his current session: such as username and log-out / log-in facilities.

The overall architecture of the INAF Catania workspace is depicted in Figure 1. The gateway is able to connect with a variety of DCIs thanks to the integrated workflow management system (see Section 2.1), the embedded ACID environment allows both on-line and off-line analysis through commonly used CTA tools (see Section 3). It is connected to the AAI to open the access to the CTA community accordingly to each user own role and/or access right.

Fig. 1: INAF Catania Workspace Architecture.

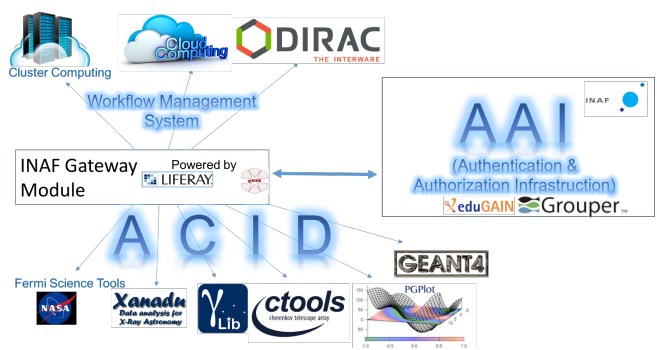

### A. Workflow Management System

Scientific workflow management systems [8] offer means to compose and distribute steps needed to perform computations for data analysis or simulations, whereas hiding details about the complex infrastructures underneath [9], [10]. More importantly, workflow descriptions capture the process of scientific experimentation, which are useful to reproduce, reuse or re-purpose these processes [11].

A plethora of mature workflow systems has evolved that support diverse concepts and languages with different strengths and focus on different modes of processing. Few workflow systems deliver the power of diverse digital resources and most of the web-based creation and editing tools either require local software installations with inherent security problems or offer incomplete functionalities. Therefore gUSE has been selected mainly because of: i) its usability via web-based user interfaces; ii) its availability, with respect to licensing terms and cost; iii) its anticipated long-term support, e.g. via an active open-source community; and iv) its ability to deal efficiently with the scales of data, computation and concurrent use required [12]. gUSE enables users convenient and easy access to distributed computing infrastructures (clusters, DIRAC INTERWARE, clouds) by providing a general purpose, workflow-oriented web-based user interface WS- PGRADE consisting of web services for the workflow management and accessing various distributed data storages.

---

[4]List of Software and tools available in the INAF CTA science gateway: http://acid.oact.inaf.it/ACID/Included_packages.html

[5]DIRAC (Distributed Infrastructure with Remote Agent Control) INTERWARE web page: http://diracgrid.org/

[6]ARC (Advence Resource Connectior) web page: http://www.nordugrid.org/arc/

[7]GLOBUS web page: https://www.globus.org/

[8]gLite web page: http://cern.ch/glite

[9]UNICORE (Uniform Interface to Computing Resources) web page https://www.unicore.eu/

[10]PBS (Portable Batch System) web page http://www.pbsworks.com/

*B. Fermi Workflow Demonstrator*

A demonstrator has been implemented following a typical Fermi analysis performed with the Fermi Science Tools[11]. Fermi Science Tools and data analysis chain is used to simulate CTA analysis. Tools and data are public and can be found at http://fermi.gsfc.nasa.gov/ssc/data. Figure 2 gives the diagram of the test case. Two analysis chains are available and should give similar results. The 2 chains (BINNED and UNBINNED) share few steps in common. These chains are standard analysis steps for CTA analysis.

Fig. 2: Fermi analysis test case.

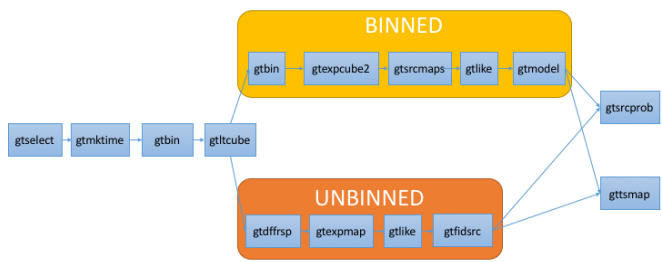

The Fermi analysis requires 3 input files: an event file in FITS format, a spacecraft file in FITS format, and the sky model in XML format.

The user has to provide also few parameters for the analysis:

- the position of the target : Right Ascension (RA in degrees) and Declination (Dec in degrees),
- the energy ranges in MeV (Minimum energy Emin and maximum energy Emax),
- the time range in MET (Mission Elapsed Time, start time and stop time),
- the radius of the region of interest to use (ROI in degrees),
- the instrument response functions (IRFS).

The output of this use case are a set of FITS files returning the processed maps.

On the science gateway two workflows have been implemented: BINNED and UNBINNED running on to the INAF Astrophysical Observatory of Catania clusters: muoni-server-02.oact.inaf.it, and acid.oact.inaf.it. The workflows have been designed to set the input datasets and the parameters to run the process into the InputSet job so that only the entry job is configured and then the parameters are passed to the other jobs automatically.

---

[11]Fermi Science Tools web page: http://fermi.gsfc.nasa.gov/ssc/data/analysis/scitools/overview.html

The separation of the Fermi processing into different jobs within the workflow allowed us to exploit the full parallelization of the computations within the configured DCIs. Finally the OutputFileSet job collects all the jobs output files and send them also to the ownCloud[12] server which is synchronized with the user account hosted into the ACID environment.

Fig. 3: BINNED workflow developed within the INAF CTA science gateway.

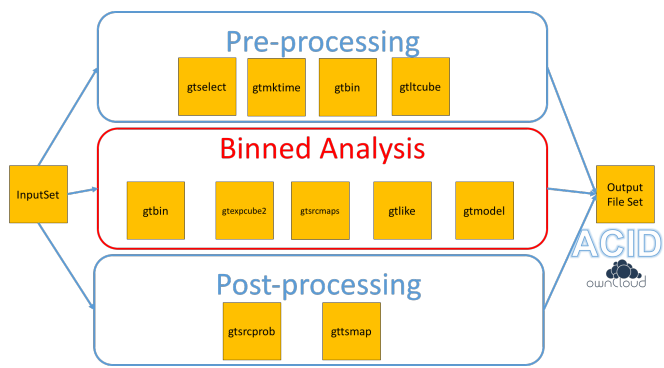

### III. ACID

The Astronomical & Physics Cloud Interactive Desktop (available at the following URL: http://acid.oact.inaf.it/ACID/Home_page.html) [13] allows to use many software packages without any installation on local desktops. Through the ACID environment the users are able to exploit the native Graphical User Interface of the available applications. A long list of astronomical and physics software suites are already available in ACID including among others: ctools & GammaLib, Fermi Science Tools, Geant4 PGPlot. Moreover it uses ownCloud to easily share data between the user device and the ACID server(s).

Fig. 4: ACID usage modes from the science gateway: on-line analysis and off-line analysis.

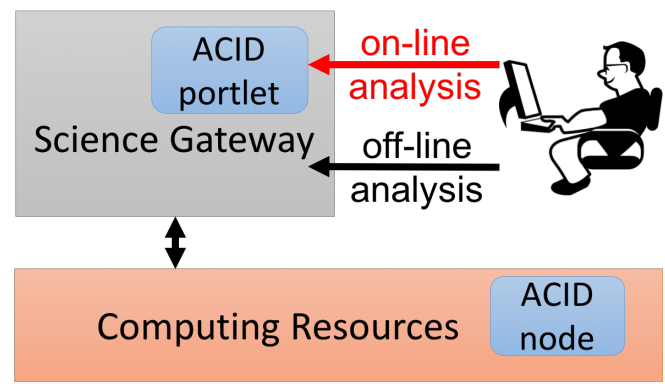

ACID is exploited by the science gateway offering two modes of usage (see Figure 4):

---

[12]Owncloud web site: http://www.owncloud.com

Fig. 5: Geographical distribution of CTA consortium members.

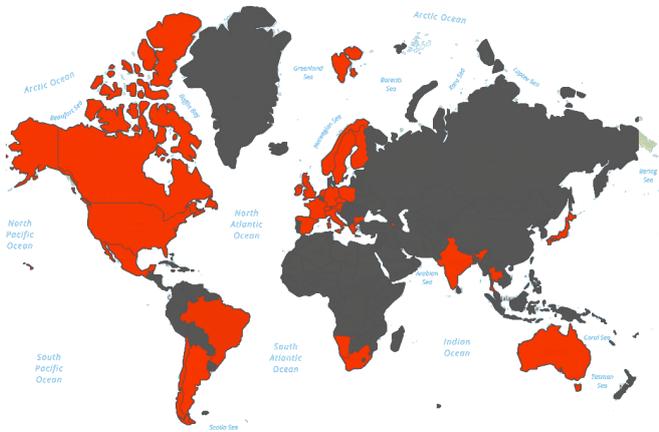

- on-line analysis: to perform interactive analysis through the Virtual Network Computing (VNC) and the native GUIs or shell environment of the CTA applications. It does not require any installation from the user since the browser loads and launches ACID as a Java Applet.
- off-line analysis: through workflow submissions to ACID employed as Distributed Computing Infrastructure (e.g. as a node of a cluster). In this case a workflow job can be configured to exploit the large number of command line-based software packages available in the ACID environment.

## IV. Authentication and Authorization Infrastructure

The CTA consortium is an experimental scientific collaboration, it consists of over 1200 members working in 32 countries from 200, mostly academic, institutes. The geographical location of consortium members, as shown in Figure 5, leads to the need of a pervasive Federated Identity Management network.

### A. Authentication

eduGAIN[13] is a service developed within the GANT network Project which is a major collaboration between European national research and education network (NREN) organisations and the European Union. eduGAIN interconnects identity federations around the world, simplifying access to content, services and resources for the global research and education community. It enables the trustworthy exchange of information related to identity, authentication and authorisation by coordinating elements of the federation technical infrastructure and providing a policy framework that controls this information exchange.

We see eduGAIN as the best approach to achieve a CTA consortium-wide authentication Infrastructure since the majority of the consortium members already belong to eduGAIN.

---

[13]eduGAIN web site:http://services.geant.net/edugain/Pages/Home.aspx

INAF is therefore running an AAI (Authentication and Authorization Infrastructure) composed by a Shibboleth[14] CTA Identity Provider and a Grouper[15] authorization solution fully compatible with the eduGAIN standards. The INAF CTA science gateway module is able to handle users authenticated by multiple federated Identity Provides, this is done by a Shibboleth Discovery Service[16]. The CTA Identity Provider acts as a cross-border/cross-domain CTA access complementing the eduGAIN identity federation and granting access to each CTA consortium user even in the case he is not (yet) member of any national federation.

Federations participating in eduGAIN adhere to a common lightweight technical and policy infrastructure and post their local federation policies so that others can learn about their registration practices and other relevant details. Each national federation already publishes a trust registry in the form of a metadata file. Each federation sends its registry to eduGAIN, except the entries that a member organization does not want to be included. eduGAIN combines all the national registries and republishes them in one large file (see Fig. 6). A national federation imports the eduGAIN consolidated international registry, merges it with our local entries and publishes them for your use.

Fig. 6: eduGAIN structure.

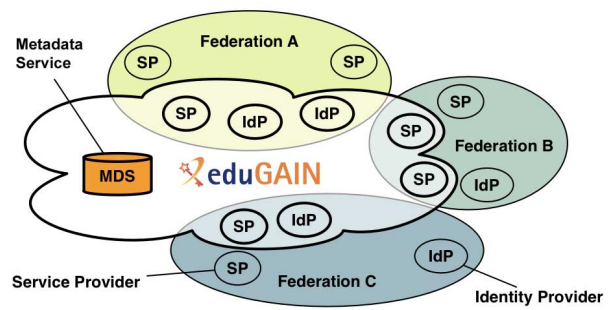

*1) Discovery Service:* In the CTA INAF science gateway the user can be authenticated by multiple IdPs (in principle each IdP of the eduGAIN inter-federation); this is currently done by using the Shibboleth Discovery Service. The Discovery Service is the process by which a service provider identifies the appropriate Identity Provider. In the case shown in this paper, the discovery service watches the federation metadata providing the user with a list identity providers. It also provides a "Search-as-you-type" selection: an effective search that guides the user in creating and reformulating her/his selection.

### B. Authorization

On top of the eduGAIN federated authentication infrastructure INAF Catania is providing a CTA authorization solution based on Grouper. Grouper is an authorization solution that keeps the membership affiliation consistent across multiple applications allowing to create and manage groups. Groups are used within each CTA application (e.g. a science gateway, the archive user interface or the project management portal) to track an individual role, or to determine which users are authorized to access the resources. If groups are managed separately in each application, keeping the membership list consistent across these services becomes very difficult.

---

[14]Shibboleth web site: https://shibboleth.net/
[15]Grouper web site: http://www.internet2.edu/products-services/trust-identity-middleware/grouper/
[16]Shibboleth Discovery Service web site: https://wiki.shibboleth.net/confluence/display/SHIB2/DiscoveryService

Grouper provides a way to define a group once and use that group across multiple applications managing it at a single point. The single point of control implies that, once a person is added or removed from a group, the group-related privileges are automatically updated in all of the collaborative applications. The current Grouper prototype proposed by INAF Catania is designed to manage and release the isMemberOf attribute and the eduPersonEntitlement attribute. These attributes are consumed by the CTA INAF science gateway and are used to authorize the access to specific web resources. These attributes are released using standard SAML assertions. Figure 7 shows the architectural diagram of the Grouper-based AAI prototype.

Fig. 7: Authentication & Authorization Infrastructure.

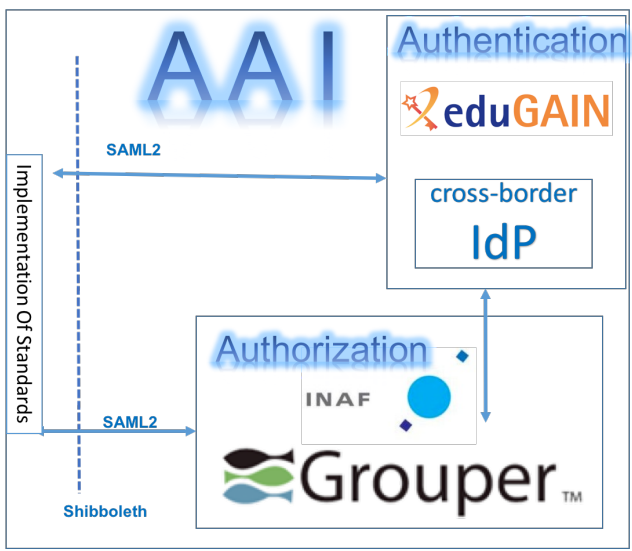

### C. Connection with the Science Gateway

To connect the INAF CTA science gateway with the proposed AAI an open-source Shibboleth Plugin available on GitHub[17] has been employed and configured within the gateway. The plugin provides an interface on the Liferay Control Panel, it allows for mapping of attributes and, thanks to the role mapping feature of Liferay it has been exploited for both authentication and authorization.

The INAF CTA science gateway has been protected with Shibboleth by running Apache HTTP[18] server in front of the Tomcat[19] servlet container. It has been configured to run on a private address and the Apache server intercepts all requests passing them to the gateway using AJP[20] (Apache JServ Protocol).

The Shibboleth Service Provider has been configured to include the attribute prefix as "AJP_", otherwise user attributes from Shibboleth could not be accessible in the gateway. It was set the AJP communication with the backend and configured Shibboleth to be "activated" for the whole gateway and required a Shibboleth session at the login.

We have so far identified two roles:

- Advanced WF User: Creates new workflows with the science gateway WorkFlow management User Interface
- WF User: Defines the workflow parameters, launches workflow processes and check results

The attribute "isMemberOf" is mapped within the gateway control panel as shown in the Figure 8 and the role permissions are set at portlet level in order to allow/deny access to specific gateway functionalities. In this case the "Advanced WF User" role will have full access to the whole workflow management system functionalities (e.g. workflow design and implementation) while the "WF User" will be enabled only to import pre-defined workflows developed by the advanced workflow users and customize them to run.

Fig. 8: Shibboleth configuration within the INAF CTA science gateway.

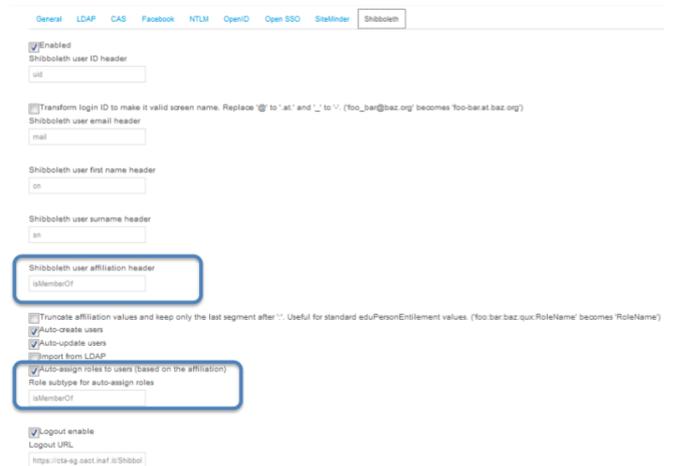

## V. RELATED WORK

Related prototyping works in the context of the CTA Science Gateway activities comprise two other modules namely:

---

[17]Liferay Shibboleth Plugin: https://github.com/ivan-novakov/liferay-shibboleth-plugin

[18]Apache HTTP web site: https://httpd.apache.org

[19]Tomcat web page: http://tomcat.apache.org

[20]AJP documentation: https://tomcat.apache.org/tomcat-7.0-doc/config/ajp.html

the Cyfronet InSilicoLab module and the Data Distiller module developed by the Observatoire de Paris. The InSilicoLab gateway supports Monte Carlo simulations performed on distributed computing infrastructures (grids) thanks to the integration with the DIRAC middleware. The Data Distiller is implemented using the Python based Django framework [21]. This prototype allows to search, retrieve and analyse high level CTA data products using the Virtual Observatory standards [22].

Regarding the AAI for CTA, apart from the one discussed in this paper, a UNITY [23] (UNIfied identiTY) management prototype is under testing. It supports multiple authentication protocols to allow integration with various consumers/clients and has the ability to outsource credentials management to a 3rd party service. The Grouper-based AAI has been preferred by the authors instead of the UNITY one because of its reliability: it is deployed at many Universities and other organizations on 4 continents. Furthermore it provides features not currently available in UNITY such as the connection with the System for Cross-domain Identity Management[24] (SCIM) or VOOT[25] (an extensible protocol for dynamic exchange of group and authorization data), bulk user import/export and user data (memberships) expiration.

## VI. CONCLUSIONS AND FUTURE WORKS

In this paper, we have introduced a workspace tailored to the requirements of the CTA community. The workspace consists of a science gateway module based on the Liferay framework endowed of a workflow management system and embedding a web-desktop environment (ACID). The workspace provides an authentication and authorization infrastructure. We described the possibilities of ACID for on-line interactive analysis and off-line processing. We presented the Fermi Workflow Demonstrator used as a test-bench for the typical CTA analyses. We highlighted the possibilities for exploiting the full job parallelization of the workflow management system and the connection with the ACID environment for cloud storage of output results. Finally we detailed the implementation and usage within the science gateway of the authentication and authorization infrastructure which guarantee the access to CTA users tuned according to his/her own role and/or access rights.

Wide adopted standards (such as SAML 2.0 and Shibboleth 2.0) and open-source technologies (such as WS-PGRADE/gUSE and Grouper) have been adopted within the proposed workspace. This aims at enlarging the developer community and improving the sustainability of the workspace during the whole CTA lifetime. The proposed solution provides an highly flexible ecosystem in order to tailor a product suitable to the present and future requirements of the CTA community.

The next steps within this work are foreseen to be focused on the integration of the proposed workspace with the other modules and services of CTA. In particular the consortium is focusing on solutions to provide messaging protocols between the different modules. Moreover we will give support for the integration with the developed AAI to the other modules.

[21] Django web site: https://www.djangoproject.com
[22] IVOA web site: http://www.ivoa.net
[23] UNITY web site: http://unity-idm.eu
[24] SCIM web site: http://www.simplecloud.info
[25] VOOT: http://openvoot.org

ACKNOWLEDGMENT

The authors would like to thank colleagues from the CTA DATA Management group, in particular Nadine Neyroud, Bruno Khelifi from LAPP (FR); Tomasz Szepieniec, Joanna Kocot, Hubert Siejkowski from Cyfronet (PL); a part of this work was developed within SCI-BUS (FP7-INFRASTRUCTURES-2011 contract 283481) project.